\documentclass[onecolumn,amsmath,amssymb,11pt]{revtex4-1}
\def\lsim{\raise0.3ex\hbox{$\;<$\kern-0.75em\raise-1.1ex\hbox{$\sim\;$}}}
\def\gsim{\raise0.3ex\hbox{$\;>$\kern-0.75em\raise-1.1ex\hbox{$\sim\;$}}}
\def\Frac#1#2{\frac{\displaystyle{#1}}{\displaystyle{#2}}}
\def\x#1#2{#1 \times #2}

 \normalsize
\newcommand{\bmat}{\left(\begin{array}}
\newcommand{\emat}{\end{array}\right)}
\newcommand{\be}{\begin{equation}}
\newcommand{\ee}{\end{equation}}
\newcommand{\bea}{\begin{eqnarray}}
\newcommand{\eea}{\end{eqnarray}}

\newcommand{\mat}[1]{\begin{pmatrix} #1 \end{pmatrix}}
\newcommand{\eq}[1]{Eq.~(\ref{#1})}
\usepackage[dvips]{graphicx}
\usepackage{graphics}
\usepackage{psfrag}
\graphicspath{{./Figures/}}

\newcommand{\n}{\nonumber\\}

\usepackage{graphics}
\usepackage{epsfig}
\usepackage{psfrag}
\begin{document}
\begin{flushright}
UMISS-HEP-2013-07\\
\end{flushright}

\title{Higgs Vacuum Stability in $B-L$ extended Standard Model}
\author{Alakabha Datta$^{1}$, A. Elsayed$^{2,3}$, S. Khalil$^{2,4}$ , A. Moursy$^{2}$ }
\vspace*{0.2cm}
\affiliation{$^1$Department of Physics and Astronomy, 108 Lewis Hall,
University of Mississippi, Oxford, MS 38677-1848, USA.\\
$^2$Center for Theoretical Physics, Zewail City for Science and Technology, 6 October City, Cairo, Egypt.\\
$^3$Department of Mathematics, Faculty of Science, Cairo
University, Giza, 12613, Egypt.\\
$^4$Department of Mathematics, Faculty of Science,  Ain Shams
University, Cairo, 11566, Egypt.}

\date{\today}

\begin{abstract}

We study vacuum stability of $B-L$ extension of the Standard Model
(SM) and its supersymmetric version. We show that the generation
of non-vanishing neutrino masses through TeV inverse seesaw
mechanism leads to a cutoff scale of SM Higgs potential stability
of order $10^5$ GeV. However, in the non-supersymmetric $B-L$
model, we find that the mixing between the SM-like Higgs and the
$B-L$ Higgs plays a crucial role in alleviating the vacuum
stability problem. We also provide the constraints of stabilizing
the Higgs potential in the supersymmetric $B-L$ model.

\end{abstract}
\maketitle

\section{Introduction}
\label{sec:intro} Recent results announced by ATLAS and CMS
experimental collaborations at the Large Hadron Collider (LHC) \cite{Aad:2012tfa,Chatrchyan:2012ufa}
confirmed the discovery of a Higgs boson with mass of order 125 GeV.
Both ATLAS and CMS have performed searches for the Higgs boson in
the following five decay channels: $H \to \gamma \gamma$, $H \to
ZZ^{(*)} \to 4 l$, and $H \to WW^{(*)} \to l \nu l \nu$, $H\to
\tau^+ \tau^-$ and $H\to b \bar{b}$, at integrated luminosities of
$5.1~ fb^{-1}$ at energy $\sqrt{s} =7$ TeV and $19.6~ fb^{-1}$ at
$\sqrt{s}=8$ TeV.

One important question is whether this scalar boson is compatible
with Standard Model (SM) predictions or it is a SM-like Higgs of an
extension of the SM. It is worth mentioning that the signal strength
of $H\to \gamma \gamma$ seems not consistent with the SM
predictions \cite{Moriond1:2013,Moriond2:2013}. It is found to be of order $1.65$
by ATLAS and about $0.78$ by CMS, while the corresponding SM
signal strength should be exactly one. In addition, it is well
known that if the SM Higgs mass is less than 130 GeV, then the quartic
Higgs self-coupling runs to negative values at high energy scales,
leading to vacuum instability at these scales~\cite{Hambye:1996wb, Djouadi:2005gi, Ellis:2009tp, Zoller:2012cv, Degrassi:2012ry, Sher:1988mj, Lindner:1988ww, Alekhin:2012py, Masina:2012tz, Sher:1993mf}. In particular, for
Higgs mass of order 125 GeV, one finds that the cutoff scale of stability for the
 SM Higgs potential  is of the order ${\cal O}(10^{9\textnormal{-}10})$ GeV.
A natural solution for this problem is to consider a possible new
physics beyond the SM that changes the running of the quartic
coupling and prevents it from running into negative values \cite{Datta:1999dw, Jiang:2013pna, Machida:2013zxa, Carena:2012mw, Anchordoqui:2012fq, Datta:2012db, Chao:2012xt, Xing:2011aa, EliasMiro:2011aa, Wingerter:2011dk}.
One can also study the issue of vacuum stability in a model independent way in an effective Lagrangian framework \cite{Datta:1996ni}. The addition of a higher dimensional operator to the Higgs potential changes the boundary condition for the quartic coupling at the scale of vacuum stability. In this work the effect of the higher dimensional operator will be neglected and only the running of the couplings will be used to determine vacuum stability.

Non-vanishing neutrino masses are now firm evidence for an
extension of the SM. One of the attractive scenarios for
accommodating the neutrino masses is the inverse seesaw mechanism,
which is based on the extension of the SM with TeV scale right
handed neutrinos with unsuppressed couplings to the SM leptons
\cite{Ma:2009gu, Khalil:2010iu, Law:2013gma, Dev:2012bd,
Gogoladze:2012jp, Dev:2012sg, Das:2012ze, BhupalDev:2012zg,
Dias:2012xp, LalAwasthi:2011aa, Abud:2011mr, Dias:2011sq,
Bazzocchi:2010dt, Park:2010by, Ibarra:2010xw, Bergstrom:2010qb,
Dev:2009aw, Hirsch:2009ra, Malinsky:2009dh, He:2009xd, Ma:2009kh,
Stephan:2009pn, Garayoa:2006xs}. In this case, one can show that
the contribution of the right-handed neutrinos has a large impact
on the Higgs quartic coupling and, similar to the top
contribution, drives it to negative values. Therefore, the SM Higgs
potential is unstable at a scale of order ${\cal
O}(10^{5\textnormal{-}6})$ GeV and the vacuum stability problem becomes more
severe. The investigation of vacuum stability within different
type of seesaw mechanisms have been explored in Refs.
\cite{Kobakhidze:2013pya, Chen:2012faa, Rodejohann:2012px,
Chakrabortty:2012np, Chun:2012jw, Khan:2012zw}.

In this article, we analyze the vacuum stability problem in simple
extensions of the SM. In particular, we focus on the $B-L$
extension of the SM with and without supersymmetry. The $B-L$
model is based on the gauge group $SU(3)_C \times SU(2)_L \times
U(1)_Y \times U(1)_{B-L}$
\cite{Marshak:1979fm,Mohapatra:1980qe,Khalil:2010zza}. It
naturally introduces three SM singlet fermions to cancel the $U(1)_{B-L}$
anomalies and account for the current experimental results of
light neutrino masses and their large mixings
\cite{Khalil:2006yi}. In addition, the extra-gauge boson and the
extra-Higgs, predicted in the $B-L$ model, have  interesting
phenomenology that can be probed at the LHC
\cite{Emam:2007dy,Abdallah:2011ew,Elsayed:2012ec,Basso:2008iv,Basso:2009gg}.
Within a supersymmetric context, it was emphasized that the three
relevant physics scales related to the supersymmetry, electroweak
and $B-L$ symmetry breaking are linked together and occur at the
TeV scale
\cite{Khalil:2007dr,Burell:2011wh,FileviezPerez:2010ek,FileviezPerez:2011kd}.
Finally, it is worth mentioning that within $B-L$ Supersymmetric
Standard Model (BLSSM) with inverse seesaw, the one-loop radiative
corrections to the lightest SM-like Higgs boson mass, due to the
right-handed neutrinos and sneutrinos, can be significant
\cite{Elsayed:2011de}, and hence the Higgs mass can be easily of order
125 GeV without pushing the SUSY spectrum to TeV scale as in MSSM.

We show that, in non-supersymmetric $B-L$ model with type-I seesaw
or inverse seesaw mechanisms, the non-vanishing mixing between the
SM and $B-L$ Higgs bosons raises the initial value of the SM-like
Higgs coupling. In addition, in this case the running of the
SM-like Higgs receives a positive contribution from the
($B-L$)-like heavy Higgs. Therefore, the Higgs self-coupling
remains positive all the way up to the GUT scale that ensures the
vacuum stability. We also analyze the vacuum stability of SM-like
Higg potential in supersymmetric $B-L$ model. The conditions
securing the stability of this potential in both flat and non-flat
directions are derived.

The paper is organized as follows. In section 2 we reappraise the
Higgs vacuum stability in the SM extended by TeV scale
right-handed neutrinos with inverse seesaw mechanism. Section 3 is
devoted for the Higgs vacuum stability in $B-L$ extension of the
SM. We show that the mixing between the SM-like Higgs and the
$B-L$ Higgs resolve the vacuum stability problem In section 4 we
analyze the vacuum stability in supersymmetric theories. In
particular, we consider the stability in MSSM and BLSSM. Finally,
we give our conclusions in section 5.

%%%%%%%%%%%%%%%%%%%%%%%%%%%%%%%%%%%%%%%%%%%%%%%%%%%%%%%%%%%%%%%%%%%%%%
%
\section{Vacuum Stability of SM Extended with TeV scale Right-Neutrinos} \label{sec:smnuR}
In this section, we analyze the impact of massive neutrinos on the
SM vacuum stability by extending the SM by right-handed neutrinos.
As known, the non-vanishing small neutrino masses can be generated
through type-I seesaw mechanism or inverse seesaw mechanism. In
type-I seesaw, one assumes that the SM lagrangian is extended as
follows:%
\be%
{\cal L}={\cal L}_{\rm{SM}}+Y_\nu \bar{l}\tilde{\Phi}\nu_R+ M
\overline{\nu}_R^c \nu_R , %
\ee%
where $\nu_R$ is a SM singlet fermion, called the right-handed
neutrino and $M$ is Majorana mass which is not restricted by the
electroweak symmetry breaking scale, so it can take any value up
to any high scale. In this case, one finds that the lightest
neutrinos get the following masses $m_\nu \sim {(Y_\nu v)^2 \over
M}$, where $v=\langle \phi \rangle$ is the electroweak VEV.
Therefore, if $M \sim {\cal O}(1)$ TeV, the light neutrino masses
can be of order electron volt, provided that $Y_\nu \sim 10^{-6}$.
In this case the contribution of the right handed neutrinos to the
Renormalization Group Equation (RGE) of the Higgs quartic coupling
is negligible, and one ends with the SM results for the Higgs
vacuum stability.

We now turn to inverse seesaw mechanism. In this case, three extra
SM singlet neutral fermions $S_i$ are required in addition to the
three right-handed neutrinos $\nu_{R_i}$ and the lagrangian in this case is given by%
\be%
{\cal L}= Y_\nu \bar{l}\tilde{\Phi}\nu_R+ M
\overline{\nu}^c_R S  + \mu_s \overline{S}^c S +h.c. %
\ee%
Thus, the neutrino mass matrix is given by
\be
\left(%
\begin{array}{ccc}
  0 & v Y_\nu& 0\\
  v Y_\nu^T & 0 & M \\
  0 & M^T & \mu_s\\
\end{array}%
\right).%
\ee%
Hence, the light neutrino masses are given by
\bea%
m_{\nu_l} &=& v^2 Y_\nu M^{-1} \mu_s (M^T)^{-1} Y_\nu^T,
\eea %
which can be of order eV, as required by the oscillation data, for
$M \sim {\cal O}(1)$ TeV if $\mu_s$ is sufficiently small, namely,
$\mu_s \lsim 10^{-7}$ GeV. In this case, the Yukawa coupling $Y_\nu$
can be of order one. Hence, the right-handed neutrino's
contribution to the RGE of the Higgs quartic coupling $\lambda$,
which is proportional to the neutrino Yukawa coupling $Y_\nu$
\cite{Luo:2002ey}, can be significant
\bea
 \frac{{\rm d} \lambda}{{\rm d }t} = \frac{1}{16\pi^2} \left[
24 \lambda^2 + 4 \lambda( 3Y_t^2+Y_\nu^2)- 2(Y_\nu^4+ 3Y_t^4)-
3\lambda (3g_2^2+ g_1^2) +\frac{9}{8}g_2^4 +\frac{3}{8}g_1^4
+\frac{3}{4}g_2^2g_1^2  \right]\, . \label{lambda}
\eea
In addition, the RGEs of top and neutrino Yukawa couplings are
given by
\bea\label{RGESMRn}
\frac{d}{dt}Y_t &=& \frac{Y_t}{16\pi ^2}\left( \frac{9}{2}Y_t^2+Y_\nu^2-8g_s^2-\frac{9}{4}g^2-\frac{17}{12}g_1^2 \right)\,\\ \nonumber
\frac{d}{dt}Y_\nu &=& \frac{Y_\nu}{16\pi ^2}\left( \frac{5}{2}Y_\nu^2+3Y_t^2-\frac{9}{4}g^2-\frac{3}{4}g_1^2 \right)\,
\eea
\begin{figure}[t!]
  \psfrag{X}[c]{$\log_{10}(Q/{\rm GeV})$}
  \psfrag{Y}[cb]{$\lambda$}
  \psfrag{LBL}[c]{$\ $}
  \psfrag{QCD1}[lb]{\tiny$m_t=172.3,\;\alpha_s=0.1191$}
  \psfrag{QCD2}[lb]{\tiny$m_t=173.2,\;\alpha_s=0.1184$}
  \psfrag{QCD3}[lb]{\tiny$m_t=174.1,\;\alpha_s=0.1177$}
\includegraphics[width=0.60\textwidth ]{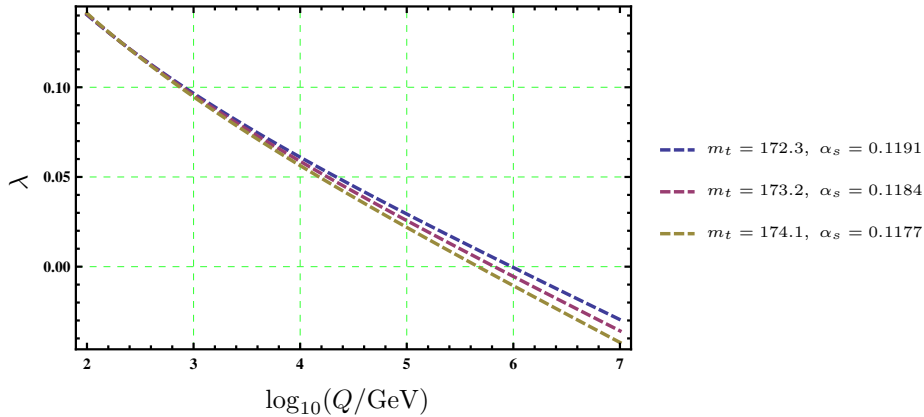}
\caption{The running of the quartic Higgs coupling for Higgs mass
$m_h=125$ GeV and $Y_\nu=0.7$ in the extended SM with right handed
neutrinos and inverse seesaw mechanism.}
  \label{smrnu}
\end{figure}

In Fig.~\ref{smrnu} we display the running of the Higss self
coupling $\lambda$ in the extended SM with right-handed neutrinos
with inverse seesaw for Higgs mass $m_h=125$ GeV. From this
figure, it is clear that the scale of Higgs vacuum stability is
reduced from $10^{9\textnormal{-}10}$ GeV in the SM to
$10^{5\textnormal{-}6}$ GeV. This can be easily understood from
the RGE (\ref{lambda}), where the neutrino Yukawa coupling $Y_\nu$
contributes to the evolution of $\lambda$, with fourth power and
negative sign, similar to the top Yukawa coupling contribution.
Therefore, one can conclude that solving the puzzle of neutrino
masses in the context of the SM gauge group with inverse seesaw
mechanism affects the Higgs vacuum stability negatively.

%%%%%%%%%%%%%%%%%%%%%%%%%%%%%%%%%%%%%%%%%%%%%%%%%%%%%%%%%%%%%%%%%%%%%%%%%%%%%%%%%
\section{Vacuum stability in $U(1)_{B-L}$ extension of the SM}
\label{B-L}

TeV scale $B-L$ extension of the SM, which is based on the gauge
group $SU(3)_C \times SU(2)_L\times U(1)_Y \times U(1)_{B-L}$ is
one of the most straightforward extensions of the SM. It permits
to introduce naturally three right-handed neutrinos, with $B-L$
charge $=-1$, due to the anomaly cancellation condition. In the
$B-L$ model with type-I seesaw mechanism \cite{Buchmuller:1991ce,
Wetterich:1981bx, Mohapatra:1980qe, Kanemura:2011mw, Pruna:2011me,
Basso:2011hn, HernandezPinto:2011wx, Coutinho:2011xb,
Montero:2011jk, Pelto:2010vq, Kanemura:2011vm, Lindner:2011it,
Ibe:2011hq, Basso:2009zz, Buchmuller:2012wn, Ishimori:2012sw,
Okada:2012np, Orikasa:2012zz, Kajiyama:2012xg, Iso:2012jn,
Khalil:2012zz, Basso:2012gz, O'Leary:2011yq, Burell:2011wh,
Khalil:2011kd, Cao:2011cp, Khalil:2011tb}, the $U(1)_{B-L}$ is
spontaneously broken by a SM singlet scalar $\chi$ with $B-L$
charge $=+2$ which acquires a VEV $v^{\prime}$. Since the kinetic
mixing term between the field strength tensors of $U(1)_Y$ and
$U(1)_{B-L}$ is allowed by gauge symmetry, the gauge-invariant
kinetic lagrangian is given by
\begin{equation}
  {\cal L} = -\Frac14 F_{\mu\nu} F^{\mu\nu} -\Frac14 F'_{\mu\nu} F'^{\mu\nu} - \Frac{\kappa}2 F_{\mu\nu} F'^{\mu\nu}.
\label{Lmixing}
\end{equation} This mixing can be absorbed by a
suitable transformation of the gauge fields that will modify the
covariant derivatives. This can be understood as follows: from
Eq.\ref{Lmixing} one can write the covariant derivative as
\begin{equation}
D_\mu = \partial_\mu - i Q^T_\phi G A_\mu,
\end{equation}
where $Q_\phi$ is a vector containing the charges of the field
$\phi$ with respect to the two abelian gauge groups, $G$ is the
gauge coupling matrix:
\begin{equation}
  G = \mat{
           g_{_{YY}} & g_{_{YB}}\\
           g_{_{BY}} & g_{_{BB}}
        },
\end{equation}
and $A_\mu$ is given, in terms of the $U(1)_Y$ and $U(1)_{B-L}$ gauge
bosons, as
\begin{equation}
  A_\mu = \mat{ A_\mu^Y \\ A_\mu^{B-L} } .
\end{equation}
One can perform an orthogonal rotation $O$ of the gauge fields
$A_\mu$, without reintroducing the kinetic mixing, such that
\begin{equation}
  Q^T_\phi G A = Q^T_\phi G (O^T O) A = Q^T_\phi \tilde{G} B,
\end{equation}
where $\tilde{G}=G O^T$ and $B=OA$. If one chooses the orthogonal
matrix $O=\mat{c_{\theta} & s_{\theta}\\-s_{\theta} & c_{\theta}}$
such that:
\begin{eqnarray}
c_\theta & = & \frac{g_{_{BB}}}{\sqrt{g_{_{BB}}^2+g_{_{BY}}^2}}, \\
s_\theta & = & \frac{g_{_{YB}}}{\sqrt{g_{_{BB}}^2+g_{_{BY}}^2}},
\end{eqnarray}
then the transformed gauge coupling matrix $\tilde{G}$ takes the form:
\begin{equation}
  \tilde{G} = \mat{
                    g_1 & 0 \\
                    \tilde{g} & g'_1
                    },
\end{equation}
where
\begin{eqnarray}
  g_1 & = & \frac{g_{_{YY}} g_{_{BB}} - g_{_{YB}} g_{_{BY}}}{\sqrt{g_{_{BB}}^2+g_{_{BY}}^2}},\\
  \tilde{g} & = & \frac{g_{_{BB}} g_{_{YB}} + g_{_{YY}} g_{_{BY}}}{\sqrt{g_{_{BB}}^2+g_{_{BY}}^2}},\\
  g'_1 & = & \sqrt{g_{_{BB}}^2+g_{_{BY}}^2}
\end{eqnarray}
Therefore, the covariant derivative takes the form:
\begin{equation}\label{covder}
  D_\mu = \cdots - i g_1 Y B_\mu - i ( \tilde{g} Y + g'_1 Y_{B-L} ) B'_\mu.
\end{equation}

The neutrino Yukawa interactions are given by%
\be%
{\cal L}^\nu_Y = Y_{\nu}\overline{l}_{L} \phi \nu_{R} + Y_N \overline{\nu}_R^c \chi  \nu_{R} + {\rm h.c.}%
\label{YukawaSS}\ee
As mentioned above, with $v' \simeq {\cal O}(1)$ TeV , the
neutrino Yukawa coupling is constrained to be $\lsim 10^{-6}$ and
hence does not affect vacuum stability of the Higgs. However, in
the $B-L$ extension of the SM with inverse seesaw, the
$U(1)_{B-L}$ symmetry is spontaneously broken by a SM singlet
scalar $\chi$ with $B-L$ charge $=-1$. Also three SM pairs of
singlet fermions $S_{1,2}^i$ with $B-L$ charge $=\mp2$,
respectively, are introduced in addition to $\nu_{R_i}$ to
implement the inverse seesaw mechanism. Note that the addition of
the extra singlet fermions $S_{1,2}$ in pairs is necessary in
order to prevent the $B-L$ triangle anomalies. In this case, the
neutrino Yukawa lagrangian is given by
\begin{equation}
{\cal L}^\nu_Y = Y_{\nu}\overline{l}_{L} \phi \nu_{R} + Y_N \overline{\nu}_R^c \chi S_2 + \mu_s \bar{S}^c_2 S_2,
\end{equation}
Therefore, after the $B-L$ and the electroweak symmetry breaking,
one finds that the neutrino mass matrix can be written as
$\bar{\psi}^c{\cal M}_{\nu} \psi$ with $\psi=(\nu_L^c ,\nu_R, S_2)$ and ${\cal M}_{\nu}$  given by %
\be {\cal M}_{\nu}=
\left(%
\begin{array}{ccc}
  0 & m_D & 0\\
  m^T_D & 0 & M_R \\
  0 & M^T_R & \mu_s\\
\end{array}%
\right), %
\label{inverse}
\ee%
where $m_D=\frac{1}{\sqrt{2}}Y_\nu v$ and $ M_R = \frac{1}{\sqrt 2}Y_{N} v' $ and $\mu_s=\frac{v'^4}{4 M^3}\lsim 10^{-7}$ GeV may be generated from non-renormalizable terms like $\bar{S}^c_{2} {\chi^\dag}^{4} S_{2}/M^3$. Thus, the light and heavy neutrino masses are given by
\begin{eqnarray}
m_{\nu_l} &=& m_D M_R^{-1} \mu_s (M_R^T)^{-1} m_D^T,\label{mnul}\\
m_{\nu_H}^2 &=& m_{\nu_{H'}}^2 = M_R^2 + m_D^2. %
\end{eqnarray}
Therefore, the light neutrino mass can be of order eV with a TeV
scale $M_R$, provided that $\mu_s$ is very small. In this case, the
Yukawa coupling $Y_\nu$ is no longer restricted to a very small
value and it can be of order one.

In both scenarios of $B-L$ extensions of the SM, with type-I
seesaw or inverse seesaw mechanism, the Higgs sector in this model
consists of one complex SM scalar doublet and one complex
SM scalar singlet with the following scalar potential
$V(\phi,\chi)$ \cite{Khalil:2006yi}%
\begin{eqnarray}
    \label{BL-potential}
       V(\phi,\chi ) = m_1^2 \vert \phi \vert^2 + m_2^2 \vert \chi\vert^2 + \lambda_1 \vert \phi \vert^4 + \lambda_2 \vert \chi\vert^4 + \lambda_3 \vert \phi \vert^2 \vert\chi\vert^2 \, .
\end{eqnarray}
As in the SM, in order to ensure non-vanishing vevs of the Higgs fields $\phi,\chi$, the squared masses $m_1^2,m_2^2$ are assumed to be negative.
In order for this potential to be stable, the coefficient matrix of the quartic terms,
\begin{equation}
  \mat{
  \lambda_1 & \frac{\lambda_3}{2}\\
  \frac{\lambda_3}{2} & \lambda_2},
\end{equation}
has to be co-positive \cite{Kannike:2012pe}. The conditions of co-positivity of such a matrix are given by
\begin{eqnarray}
\label{inf_limitated}
\lambda_1 , \lambda_2 & > & 0,\\
 \frac{\lambda_3}{2} + \sqrt{\lambda_1 \lambda_2} & > & 0~.
\end{eqnarray}
The $U(1)_{B-L}$  and the electroweak gauge symmetry are broken by
the non-zero vevs: $\left< \chi \right> = v'/\sqrt{2}$ and $\left
< \phi \right> = v/\sqrt{2}$, where $v$ and $v'$  satisfy
the following minimization conditions:
\begin{eqnarray}\label{min1}
v^2 = \frac{-\lambda_2 m_1^2 + \frac{\lambda_3}{2}m_2^2}{\lambda
_1 \lambda _2 - \frac{\lambda _3^{\phantom{o}2}}{4}} \, ,~~~~~~~~
v'^2 = \frac{-\lambda_1 m_2^2 + \frac{\lambda_3}{2}m_1^2}{\lambda
_1 \lambda _2 - \frac{\lambda _3^{\phantom{o}2}}{4}} \, .
\end{eqnarray}

The mixing between the two neutral Higgs scalars leads to the mass
eigenstates fields $h$ and $H$, which are defined in terms of
$\phi^0$ and $\chi$. The physical mass eigenstates fields $h$ and $H$ are given by%
\begin{equation}\label{scalar_orthog}
\left( \begin{array}{c} h\\H\end{array}\right) = \left(
\begin{array}{cc} \cos{\theta}&-\sin{\theta}\\
\sin{\theta}&\cos{\theta}
    \end{array}\right) \left( \begin{array}{c} \phi^0\\ \chi\end{array}\right) \, ,
\end{equation}
where the mixing angel $\theta$ is given by
\begin{eqnarray}\label{tan2th}
\tan{2\theta} = \frac{\lambda _3 vv'}{\lambda _1 v^2 - \lambda _2
v'^2} \, .
\end{eqnarray}
The range of the mixing angle $\theta$ can be:
$-\frac{\pi}{2}\leq \theta \leq \frac{\pi}{2}$. Also, the masses
of light and heavy Higgs particles are given by
\begin{eqnarray}\label{mh1}
m^2_{h,H} = \lambda _1 v^2 + \lambda _2 v'^2 \mp \sqrt{(\lambda _1
v^2 - \lambda _2 v'^2)^2 + (\lambda _3 vv')^2} \, ,
\end{eqnarray}
From the above expressions, one can easily express the scalar
potential parameters: $\lambda _1$, $\lambda _2$ and $\lambda _3$
in terms of the physical quantities: $m_{h}$, $m_{H}$ and
$\sin{2\theta}$ as follows \cite{Basso:2010jm}
\begin{eqnarray}\label{smixing}\nonumber
\lambda _1 &=& \frac{m_{h}^2}{4v^2}(1+\cos{2\theta}) + \frac{m_{H}^2}{4v^2}(1-\cos{2\theta}) ,\\
\lambda _2 &=& \frac{m_{h}^2}{4v'^2}(1-\cos{2\theta}) + \frac{m_{H}^2}{4v'^2}(1+\cos{2\theta}),\\
\lambda _3 &=& \sin{2\theta} \left( \frac{m_{H}^2-m_{h}^2}{2vv'}
\right)\nonumber.
\end{eqnarray}
From these equations, one notices that the initial condition of
the SM-like Higgs quartic coupling, $\lambda_1$, at the electroweak
scale can be different from that in the SM. This, as we will
 see, can have an important impact on the evolution of
this coupling and Higgs vacuum stability.

The RGEs of the scalar couplings: $\lambda_1$, $\lambda_2$ and
$\lambda_3$ in the context of $B-L$ extension of the SM, are given
by \cite{Basso:2010jm}
\begin{eqnarray}\label{RGE_m}
\frac{d \lambda_1}{dt} &=&
\frac{1}{16\pi^2}\left(24\lambda_1^2+\lambda_3^2 + 4\lambda_1(
3Y_t^2+Y_\nu^2)- 2(Y_\nu^4+ 3Y_t^4) +\frac{9}{8} g_2^4
+\frac{3}{8} g_1^4 +\frac{3}{4}g_2^2g_1^2
+\frac{3}{4}g_2^2\widetilde{g}^2 \right.\nonumber\\
&+&
\left.\frac{3}{4}g_1^2\widetilde{g}^2+\frac{3}{8}\widetilde{g}^4
-9\lambda _1 g_2^2-3\lambda _1 g_1^2-3\lambda _1 \widetilde{g}^2
    \right),\label{RGE_lamda1}\\
\frac{d \lambda _2}{dt} &=&  \frac{1}{8\pi ^2}\left( 10\lambda
_2^2+\lambda _3^2-\frac{1}{2}Tr\left[ (Y_N)^4\right] +48
g_1^{'4}+4\lambda_2 Tr\left[ (Y_N)^2\right] -24\lambda_2g_1^{'2} \right),\label{RGE_lamda2}\\
\frac{d \lambda_3}{dt} &=&  \frac{\lambda_3}{8\pi ^2}\left(
6\lambda_1+4\lambda_2+2\lambda_3+3Y_t^2-\frac{9}{4}g_2^2-\frac{3}{4}g_1^2-\frac{3}{4}\widetilde{g}^2
+2Tr\left[ (Y_N)^2\right] - 12 g_1^{'2} + 6\frac{\widetilde{g}^2
g_1^{'2}}{\lambda _3}\right),  \label{RGE_lamda3}
\end{eqnarray}
where $\widetilde{g} $ and $g_1^{'} $  are the gauge couplings of
the $U(1)$'s mixing and U(1)$_{B-L}$ as defined in
Eq.~\ref{covder}. $Y_N$ is the Yukawa coupling defined in
Eq.~\ref{YukawaSS}. The scalar couplings $\lambda_1,\ \lambda_2$
and $\lambda_3$ are defined in Eq.~\ref{BL-potential}. For
completeness, we give also the RGEs of $g_1'$ and $\widetilde{g}$,
which can be written as \cite{Basso:2010jm}
\begin{eqnarray}\label{RGE_g1pSS}
\frac{d g_1'}{dt} &=& \frac{1}{16\pi ^2}\left[12 g_1'^3 + {32\over
3} g_1'^2\widetilde g+{41\over 6}g_1'\widetilde g^2 \right] \, ,
\\ \label{RGE_g_tilde} \frac{d \widetilde g}{dt} &=&
\frac{1}{16\pi ^2}\left[{41\over 6}\widetilde{g}\,(\widetilde
g^2+2g_1^2)+{32\over 3}g_1'(\widetilde{g}^2+g_1^2)+ 12
g_1'^2\widetilde g \right].
\end{eqnarray}
The RGEs of the gauge couplings, $g_3\, , g_2$ and $g_1$ remain
intact. Finally, the RGEs of the Yukawa couplings $Y_t$, $Y_\nu$ and
$Y_N$ are as follows \cite{Basso:2010jm}
\bea\label{RGE_yuk_top} \frac{d Y_t}{dt} &=& \frac{Y_t}{16\pi
^2}\left(
\frac{9}{2}Y_t^2-8g_3^2-\frac{9}{4}g_2^2-\frac{17}{12}g_1^2-\frac{17}{12}\widetilde{g}^2
-\frac{2}{3}g_1'^{2}-\frac{5}{3}\widetilde{g} g'_1 \right)\, ,\\
\label{RGE_yuk_nu}
\frac{d Y_\nu}{dt} &=& \frac{Y_\nu}{16\pi ^2}\left( \frac{5}{2}Y_\nu^2 + 3Y_t^2-\frac{9}{4}g_2^2-\frac{3}{4}g_1^2 -6 g_1'^2\right)\, \\
\label{RGE_nu_r_maj} \frac{dY_{_{N_i}}}{dt} &=&
\frac{Y_{_{N_i}}}{16\pi ^2}\left( 4(Y_{_{N_i}})^2+2Tr\big[
(Y_{_N})^2\big] -6 g_1'^2 \right)\, , \qquad (i=1\dots 3)\, , \eea
where, we consider the basis of real and diagonal  $Y_N$, i.e.
$Y_N \equiv \mbox{diag}\, (Y_{N_1},Y_{N_2},Y_{N_3})$. It is worth noting
that within inverse seesaw, the RGE of $B-L$ couplings $g_1'$ and
$\widetilde g$ are slightly modified, due to the impact of the two
fermions $S_{1,2}$, which are charged under $B-L$. They take the
form:
\begin{eqnarray}\label{RGE_g1pIS}
\frac{d g_1'}{dt} &=& \frac{1}{16\pi ^2}\left[27 g_1'^3 + {32\over
3} g_1'^2\widetilde g+{41\over 6}g_1'\widetilde g^2 \right] \, ,
\\ \label{RGE_g_tilde} \frac{d \widetilde g}{dt} &=&
\frac{1}{16\pi ^2}\left[{41\over 6}\widetilde{g}\,(\widetilde
g^2+2g_1^2)+{32\over 3}g_1'(\widetilde{g}^2+g_1^2)+ 27
g_1'^2\widetilde g \right]\, .
\end{eqnarray}

From Eq.(\ref{RGE_lamda1}) of the RGE of the coupling $\lambda_1$,
we notice that  the mixing parameter $\lambda_3$ contributes
positively to the evolution of $\lambda_1$, unlike the top Yukawa
and neutrino Yukawa couplings. Note that the evolution of
$\lambda_3$ (and also the running of $\lambda_1$) is enhanced by
the positive effect of the self-coupling of $B-L$ heavy Higgs,
$\lambda_2$. Therefore, with non-negligible $\lambda_3$, the scale
of Higgs vacuum stability can be pushed to higher values. In case
of inverse seesaw, where $Y_\nu \sim {\cal O}(1)$, a larger mixing
parameter is required to overcome the effects of both the top and
neutrino Yukawa couplings that pull the stability scale down.
Note, since the Higgs scalar is not charged under U(1)$_{B-L}$,
the running of $\lambda_1$ has no dependence on $g'_1$. The only
extra gauge contribution to $d\lambda_1/dt$ is due to the small
gauge mixing $\widetilde{g}$, which leads to a negligible effect.

\begin{figure}[t!]
  \psfrag{X}[c]{$\log_{10}(Q/{\rm GeV})$}
  \psfrag{Y}[cb]{$\lambda_1$}
  \psfrag{LBL}[c]{$\ $}
  \psfrag{theta1}[lb]{\tiny$\theta=0.2$}
  \psfrag{theta2}[lb]{\tiny$\theta=0.1$}
  \psfrag{theta3}[lb]{\tiny$\theta=0$}
  \includegraphics[width=0.45\textwidth ]{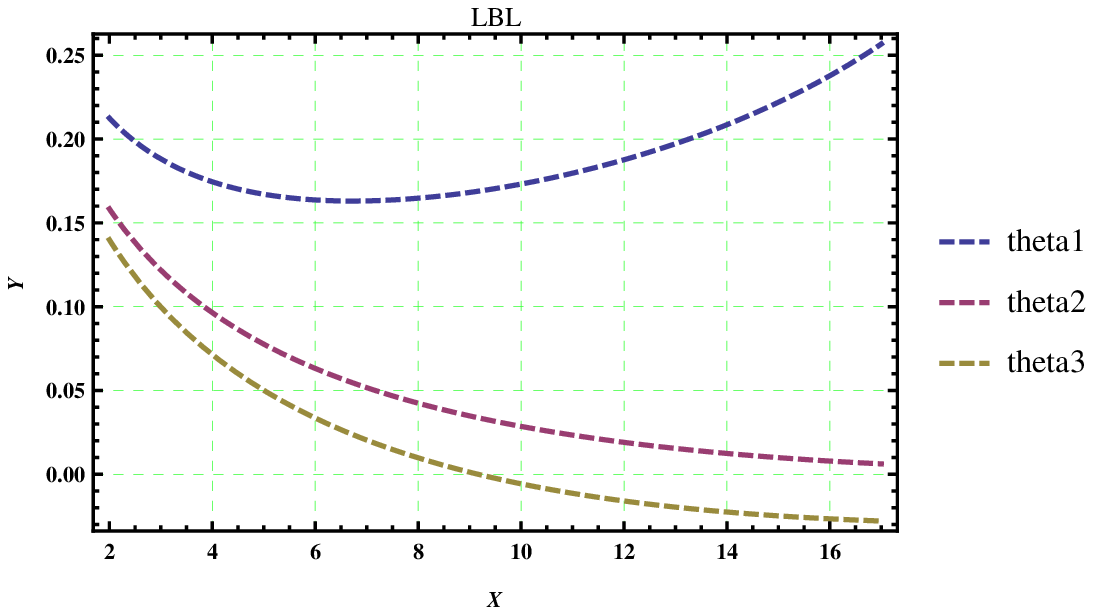} ~~~~~~~~~~~
  \psfrag{Y}[cb]{$\lambda_3+2\sqrt{\lambda_1\lambda_2}$}
  \includegraphics[width=0.45\textwidth ]{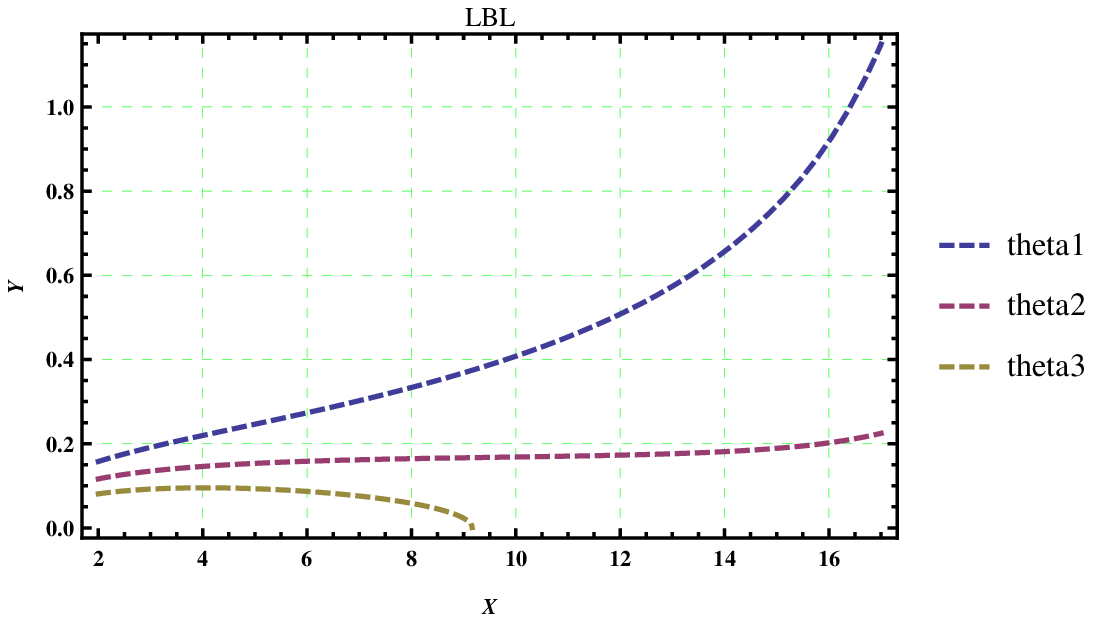}
  \caption{(Left panel): The RG running of the quartic coupling $\lambda_1$ in the $B-L$ extension of the SM with type-I seesaw, for three values of the scalar mixing angle $\theta$ for SM-like Higgs mass $m_{h}=125$ GeV. (Right panel): The evolution of the second stability condition, $\lambda_3+2\sqrt{\lambda_1\lambda_2}$, up to the GUT scale.}
  \label{BLmixss1}
\end{figure}

As emphasized, the parameter that is responsible for the scalar
mixing $\lambda_3$ is expressible in terms of the physical
quantities $m_{h}$, which is fixed by the detected Higgs
mass$~125$ GeV and the heavy Higgs mass $m_{H}$ and the mixing
angel $\theta$. In Fig. \ref{BLmixss1} we show the running, up to
the GUT scale, for the quartic couplings $\lambda_1$ and the
condition of bounded from below: $\lambda_3+2\sqrt{\lambda_1
\lambda_2}$ in the $B-L$ extension of the SM with type-I seesaw.
It is worth noting that $\lambda_2$ is unconditionally positive as
can be seen from its RG equation (\ref{RGE_lamda2}). In these
plots, we consider three values of the Higgs mixing angle: $\theta
=0, 0.1,$ and $0.2$. Also we fix the SM-like Higgs mass with 125
GeV and the heavy Higgs mass $m_H=500$~GeV. As can be seen from
this figure, at $\theta=0$ where there is no mixing between the SM
Higgs and $B-L$ Higgs, the running of $\lambda_1$ coincides with
that of the SM. Hence one again finds that the Higgs potential
becomes unstable at an energy scale $\gsim 10^{9\textnormal{-}10}$
GeV. With non-vanishing $\theta$ one finds that $\lambda_1$ gets
initial values at electroweak scale larger than its value in the
SM and also its scale dependence becomes rather different.
Therefore in this case one finds that it is quite plausible, with
not very large mixing, to keep $\lambda_1$ and also
$\lambda_3+2\sqrt{\lambda_1 \lambda_2}$ positive up to the GUT
scale, and hence the Higgs vacuum stability is accomplished.

\begin{figure}[t!]
  \psfrag{X}[c]{$\log_{10}(Q/{\rm GeV})$}
  \psfrag{Y}[cb]{$\lambda_1$}
  \psfrag{LBL}[c]{$\ $}
  \psfrag{theta1}[lb]{\tiny$\theta=0.25$}
  \psfrag{theta2}[lb]{\tiny$\theta=0.21$}
  \psfrag{theta3}[lb]{\tiny$\theta=0$}
  \includegraphics[width=0.45\textwidth ]{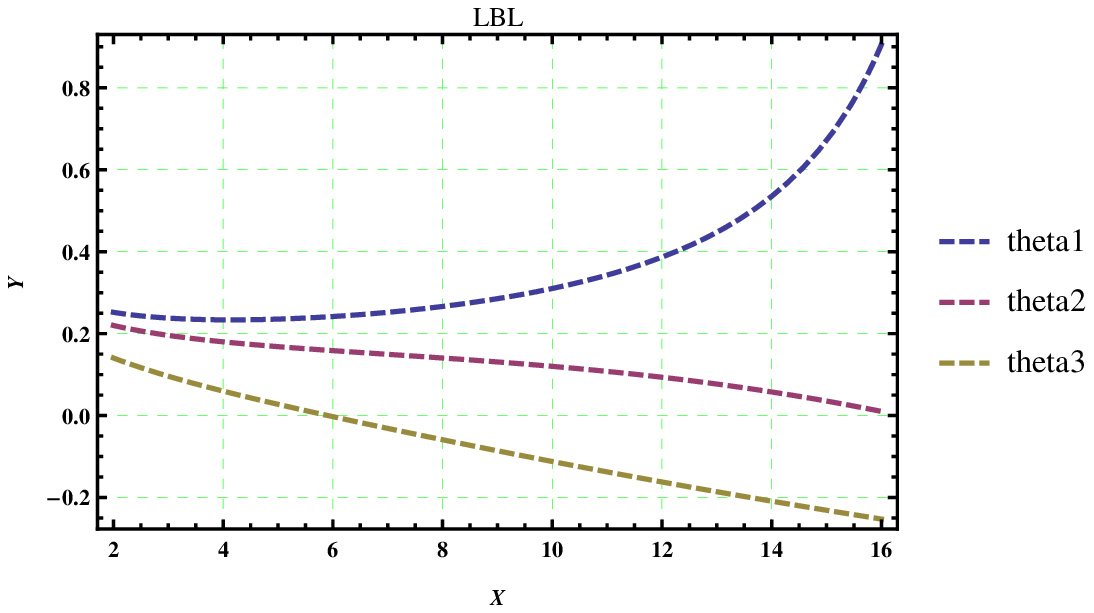} ~~~~~~~~~~~
  \psfrag{Y}[cb]{$\lambda_3+2\sqrt{\lambda_1\lambda_2}$}
  \includegraphics[width=0.45\textwidth ]{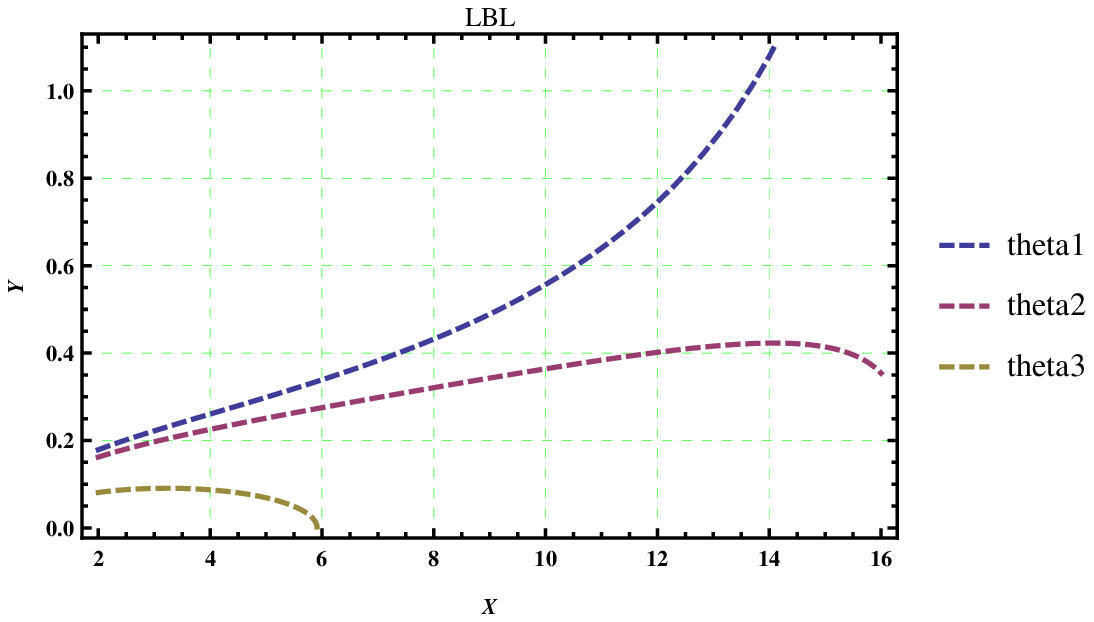}
  \caption{(Left panel): The running of the quartic couplings $\lambda_1$ in the $B-L$ extension of the SM with inverse seesaw, for three values of scalar mixing angle $\theta$ for  SM-like Higgs mass $m_{h}=125$ GeV. (Right panel): The evolution of the second stability condition, $\lambda_3+2\sqrt{\lambda_1\lambda_2}$, up to the GUT scale.}
\label{BLmixis1}
\end{figure}

Similarly, in Fig. \ref{BLmixis1} we display the running of
$\lambda_1$ and $\lambda_3+2\sqrt{\lambda_1 \lambda_2}$ in $B-L$
extension of the SM with inverse seesaw, for $\theta =0, 0.21,$
and $0.25$, $m_h=125$ GeV, $m_H =500$ GeV and $Y_\nu = 0.7$. It is
clear that with $\theta=0$, we get the non $B-L$ limit for the
instability of the Higgs potential, where both $\lambda_1$ and
$\lambda_3+2\sqrt{\lambda_1 \lambda_2}$ become negative at $\sim
10^{5\textnormal{-}6}$ GeV. Also, we find that for $\theta \gsim
0.21$, the Higgs vacuum stability is achieved up to the GUT scale.

%

%
%%%%%%%%%%%%%%%%%%%%%%%%%%%%%%%%%%%%%%%%%%%%%%%%%%%%%%%%%%%%%%%%%%%%%%%%%%%%%%%%%%%%%%%%%%%%%%%%%
\section{Vacuum Stability in Supersymmetric Extensions of the SM}
\label{MSSM}

In this section we analyze the Higgs vacuum stability in
supersymmetric extensions of the SM. We start with the MSSM, which
is the most widely studied SUSY model. The MSSM is based on the
same gauge group of the SM, i.e, $SU(3)_C \times SU(2)_L \times
U(1)_Y$, with the following superpotential
\begin{equation}
W = Y_u Q_L U^c_L H_2 + Y_d Q_L D_L^c H_1 + Y_e L_L E_L^c H_1 + \mu H_1 H_2.
\label{Superpot}
\end{equation}
In MSSM, two Higgs doublet superfields are required for the
Higgsino anomalies to cancel among themselves. From the
superpotential one can determine the scalar potential. Thus, the
potential for the neutral Higgs fields can be written
\begin{equation}
V(H_1,H_2) = m_1^2 \; H_1^2 + m_2^2 \; H_2^2 - 2 m_3^2 \; H_1 H_2 + \frac{g^2 + g'^2}{8} \left( H_1^2 - H_2^2 \right)^2,%
\label{scalar_potential1}
\end{equation}
where the masses $m^2_i$ are given in terms of the soft SUSY
breaking terms: $m^2_{H_i}$, $B$ and the $\mu$ parameter as
follows:
\begin{equation}
  m^2_i = m^2_{H_i} + \vert \mu\vert ^2 , \hspace{0.5cm} m_3^2 =  B \mu.
  \label{m12}
\end{equation}
This potential is the SUSY version of the Higgs potential which
induces $SU(2)_L \times U(1)_Y$ breaking in the SM, where the
usual self-coupling constant is replaced by the squared gauge
couplings.

In order to study the stability of the MSSM Higgs potential, one
should consider the following two cases: $(i)$ Flat direction,
where $H_1 = H_2 =: H$. $(ii)$ Non-flat directions. In the flat
direction,  the quartic terms vanish and the potential takes
the simple form:
\begin{equation}
  V(H)= (m^2_1 + m^2_2 - 2 m^2_3) H^2,
\end{equation}
which is stable only if the coefficient $(m^2_1 + m^2_2 - 2
m^2_3)$ is non-negative. This is the well known condition for
avoiding the unboundedness of MSSM potential from below.

On the other hand, on non-flat directions the quartic terms in
\eq{scalar_potential1} are non-vanishing and dominate the
potential for large value of the scalar fields $H_{1,2}$. Thus,
the stability is unconditionally guaranteed because the quartic
coupling $(g^2+g'^2)/8$ is always positive. Therefore, one
concludes that the MSSM Higgs potential is identically stable at
any direction except the flat one, which requires the following
condition:
\begin{equation}
  m^2_1 + m^2_2 \geq 2 m^2_3,%
  \label{MSSM-stability-condition}
\end{equation}

Now we turn to the supersymmetric $B-L$ extension of the SM
(BLSSM). The minimal version of BLSSM is based on the gauge group
$SU(3)_C \times SU(2)_L \times U(1)_Y \times U(1)_{B-L}$, with
particle content that includes the following fields in addition to
those of the MSSM: three chiral right-handed superfields ($N_i$),
the vector superfield necessary to gauge the $U(1)_{B-L}$
($Z_{B-L}$), and two chiral SM-singlet Higgs superfields
($\chi_1$, $\chi_2$ with $B-L$ charges $Y_{B-L}=-2$ and
$Y_{B-L}=+2$, respectively). As in the MSSM, the introduction of a
second Higgs singlet ($\chi_2$) is necessary in order to cancel
the $U(1)_{B-L}$ anomalies produced by the fermionic member of the
first Higgs superfield ($\chi_1$). The $Y_{B-L}$ for quark and
lepton superfields are assigned in the usual way.

The interactions between the Higgs and matter superfields are
described by the superpotential %
\begin{eqnarray}%
W &=& (Y_U)_{ij} Q_i  H_2 U^c_j + (Y_D)_{ij} Q_i  H_1 D_j^c +
(Y_L)_{ij} L_i H_1 E_j^c + (Y_\nu)_{ij} L_i H_2 N_j^c \nonumber\\
&+& (Y_N)_{ij} N^c_i N^c_j \chi_1 + \mu H_1 H_2
+ \mu^\prime \chi_1 \chi_2.%
\label{superpot}%
\end{eqnarray}

Therefor, the BLSSM Higgs potential is given by
\begin{eqnarray}
  V(H_1,H_2,\chi_1,\chi_2)
  & = & m_1^2 \; H_1^2 + m_2^2 \; H_2^2 - 2 m_3^2 \; H_1 H_2 + \mu_1^2 \; \chi_1^2 + \mu_2^2 \; \chi_2^2 - 2 \mu _3^2 \; \chi_1 \chi_2\n
  &   & +\; \frac{ g^2 + g_{_{YY}}^2 + g_{_{YB}}^2 }{8} \left(H_1^2 - H_2^2 \right)^2 + \frac{ g_{_{BB}}^2 + g_{_{BY}}^2 }{2}
  \left(\chi_1^2 - \chi_2^2 \right)^2\n
  &   & +\; \frac{ g_{_{BB}} g_{_{YB}} + g_{_{BY}} g_{_{YY}} }{2} \left(H_1^2 - H_2^2 \right)\left (\chi _1^2 - \chi _2^2 \right)
  \label{BLSSM-Higgs-Potential}
\end{eqnarray}
where
\begin{equation}
  m^2_i = m^2_{H_i} + \vert \mu\vert ^2 , \hspace{0.5cm} \mu^2_i = m^2_{\chi_i} + \vert \mu'\vert ^2 , \hspace{0.5cm} m_3^2 \; (\mu_3^2) =  B \mu \; (B \mu').
  \label{m12}
\end{equation}

Similar to the MSSM, in order to study the stability of this
potential, one should consider the two cases of flat direction, in
which $H_1=H_2=:H \ \&\ \chi_1=\chi_2=:\chi$, and the other non-flat
directions. In the flat direction, all the quartic terms vanish,
and the potential turns to the simple form:
\begin{equation}
  V(H,\chi)= (m^2_1 + m^2_2 - 2 m^2_3) H^2 + (\mu^2_1 + \mu^2_2 - 2 \mu^2_3) \chi^2,
\end{equation}
which is stable under the conditions
\begin{eqnarray}
  \label{BLSSM-stability-condition-flat1}
  m^2_1 + m^2_2 & \geq & 2 m^2_3,\\
  \mu^2_1 + \mu^2_2 & \geq & 2 \mu^2_3.
  \label{BLSSM-stability-condition-flat2}
\end{eqnarray}

On the other hand, the quartic terms are non-vanishing in the
other directions and they dominate the quadratic terms. Thus, the
stability is guaranteed only if the matrix of quartic terms,
\begin{equation}
  \left(
    \begin{array}{cccc}
     \frac{g^2+g_{_{YB}}^2+g_{_{YY}}^2}{8} & -\frac{g^2+g_{_{YB}}^2+g_{_{YY}}^2}{8} &
     \frac{g_{_{BB}} g_{_{YB}}+g_{_{BY}} g_{_{YY}}}{4} & -\frac{g_{_{BB}} g_{_{YB}}+g_{_{BY}} g_{_{YY}}}{4} \\
     -\frac{g^2+g_{_{YB}}^2+g_{_{YY}}^2}{8} & \frac{g^2+g_{_{YB}}^2+g_{_{YY}}^2}{8} & -\frac{g_{_{BB}} g_{_{YB}}+g_{_{BY}} g_{_{YY}}}{4} &
     \frac{g_{_{BB}} g_{_{YB}}+g_{_{BY}} g_{_{YY}}}{4} \\
     \frac{g_{_{BB}} g_{_{YB}}+g_{_{BY}} g_{_{YY}}}{4} & -\frac{g_{_{BB}} g_{_{YB}}+g_{_{BY}} g_{_{YY}}}{4} & \frac{g_{_{BB}}^2+g_{_{BY}}^2}{2} &
     -\frac{g_{_{BB}}^2+g_{_{BY}}^2}{2} \\
     -\frac{g_{_{BB}} g_{_{YB}}+g_{_{BY}} g_{_{YY}}}{4} & \frac{g_{_{BB}} g_{_{YB}}+g_{_{BY}} g_{_{YY}}}{4} &
     -\frac{g_{_{BB}}^2+g_{_{BY}}^2}{2} & \frac{g_{_{BB}}^2+g_{_{BY}}^2}{2}
    \end{array}
  \right)
  \label{quartic-BL-matrix}
\end{equation}
is co-positive. Applying the co-positivity criteria of a $\x44$
matrix \cite{Ping1993109} (See appendix
\ref{copositivity-Appendix} for brief review) implies that the
condition:
\begin{equation}
  g^2(g_{_{BB}}^2 + g_{_{BY}}^2) + g_{_{YY}}^2 g_{_{BB}}^2 + g_{_{YB}}^2 g_{_{BY}}^2 \geq 2 g_{_{YY}} g_{_{BB}} g_{_{YB}} g_{_{BY}}
  \label{BLSSM-stability-condition-non-flat}
\end{equation}
should be satisfied in order for the potential in
\eq{BLSSM-Higgs-Potential} to be stable in the non-flat direction.
It is worth noting that, in the case of no gauge mixing
($g_{YB}=0=g_{BY}$), the condition
(\ref{BLSSM-stability-condition-non-flat}) is automatically
satisfied. In this regard, the BLSSM Higgs potential is stable if
and only if the conditions in Eqs.
(\ref{BLSSM-stability-condition-flat1}),
(\ref{BLSSM-stability-condition-flat2}) and
(\ref{BLSSM-stability-condition-non-flat}) are satisfied.
\begin{figure}[t!]
    \psfrag{X}[c]{$\log_{10}(Q)$}
    \psfrag{Y}[b]{$R$}
    \psfrag{g1}{$g_{_{BB}}=0.1$}
    \psfrag{g2}{$g_{_{BB}}=0.2$}
    \psfrag{g3}{$g_{_{BB}}=0.3$}
    \psfrag{g4}{$g_{_{BB}}=0.4$}
    \psfrag{LBL}[bc]{$\ $}
    \includegraphics[width=0.5\textwidth]{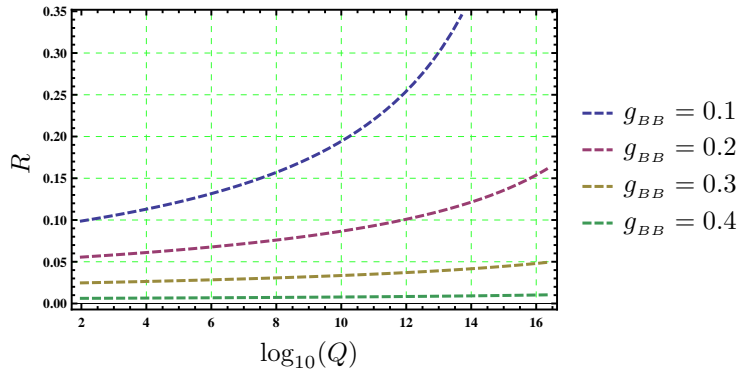}
    \caption{Running of the BLSSM condition $R\equiv g^2(g_{_{BB}}^2 + g_{_{BY}}^2) + g_{_{YY}}^2 g_{_{BB}}^2 + g_{_{YB}}^2 g_{_{BY}}^2 - 2g_{_{YY}}
    g_{_{BB}} g_{_{YB}} g_{_{BY}}$ for different initial values of $g_{_{BB}}$ at the EW-scale, fixing the initial mixing parameters
    $g_{_{YB}}\,\&\,g_{_{BY}}$ to be zero at the EW-scale.}
    \label{BLSSM-Stability-Test}
\end{figure}

In Fig.~\ref{BLSSM-Stability-Test}, we present the running of the
BLSSM stability indicator $R \equiv g^2(g_{_{BB}}^2 + g_{_{BY}}^2)
+ g_{_{YY}}^2 g_{_{BB}}^2 + g_{_{YB}}^2 g_{_{BY}}^2 - 2g_{_{YY}}
g_{_{BB}} g_{_{YB}} g_{_{BY}}$ fixing the values of the MSSM gauge
coupling at the EW-scale by its known values, and fixing the
mixing parameters $g_{_{YB}}\,\&\,g_{_{BY}}$ to be zero at the
EW-scale and varying the values of the free $g_{_{BB}}$. It is
clear that the stability indicator $R$ is always positive for any
value of $g_{_{BB}}$ which means that  no theoretical bounds can
be put on the $g_{_{BB}}$ from the stability conditions. It is
worth mentioning that the situation does not change when we relax
the conditions on the mixing gauge couplings,
$g_{_{YB}}(EW)=0=g_{_{BY}}(EW)$, by allowing nonzero values less
than $10^{-3}$\cite{Carena:2004xs}.

%%%%%%%%%%%%%%%%%%%%%%%%%%%%%%%%%%%%%%%%%%%%%%%%%%%%%%%%%%%%%%%%%%%%%%%%%%%%%%%%%%%%%%%%%%%%%%%%%%%
\section*{Conclusions}
In this paper we have analyzed the Higgs vacuum stability problem
in the $B-L$ extension of the SM and also in the MSSM. We have shown
that within the context of the inverse seesaw mechanism, which is an elegant TeV
scale mechanism for generating the neutrino masses, the Higgs
vacuum stability is affected negatively, and the cutoff scale for vacuum instability is
reduced from $10^{10}$ GeV in the SM to $10^5$ GeV. We emphasized
that the mixing between the SM-like Higgs and the $(B-L)$-like
Higgs resolves this problem due to the following reasons: $(i)$
Possible enhancement of the initial value of the SM-like Higgs
self-coupling. $(ii)$ The positive contribution of the $(B-L)$
Higgs coupling to the running of the SM-like Higgs self coupling.

We also studied the stability conditions in the supersymmetric $B-L$
model. We showed, similar to the MSSM in Higgs flat
directions,  the requirement of vacuum stability imposed constraints on the Higgs
masses. In the non-flat directions, the stability of the Higgs
potential lead to a constraint on the gauge couplings, which is
automatically satisfied if there is no kinetic mixing between
$U(1)_Y$ and $U(1)_{B-L}$.

%%%%%%%%%%%%%%%%%%%%%%%%%%%%%%%%%%%%%%%%%%%%%%%%%%%%%%%%%%%%%%%%%%%%%%%%%%%%%%%%%%%%%%%%%%%%%%%%%%%
\section*{Acknowledgments}

This work is partially supported by the ICTP project AC-80.
The work of A.D. was  supported by the US-Egypt Joint Board on Scientific
and Technological Co-operation award (Project ID: 1855) administered by the
US Department of Agriculture.
%%%%%%%%%%%%%%%%%%%%%%%%%%%%%%%%%%%%%%%%%%%%%%%%%%%%%%%%%%%%%%%%%%%%%%%%%%%%%%%%
\appendix
%\section{SM RGEs}
%%%%%%%%%%%%%%%%%%%%%%%%%%%%%%%%%%%%%%%%%%%%%%%%%%%%%%%%%%%%%%%%%%%%%%%%%%%%%%%%
%\section{U(1)$_{B-L}$ RGEs}\label{BL_RGE}

%%%%%%%%%%%%%%%%%%%%%%%%%%%%%%%%%%%%%%%%%%%%%%%%%%%%%%%%%%%%%%%%%%%%%%%%%%%%%%%%
\section{Co-positivity of Order Four Matrices}
\label{copositivity-Appendix}

The co-positivity of a square matrix can be tested through  some conditions that depends only on the dimension of the matrix as well as the signs of its elements. Such a subject is too lengthy to be presented here as a whole. Thus, we shall present the co-positivity conditions of only one class of $\x44$ matrices to-which the matrix in \eq{quartic-BL-matrix} belongs.

Consider a symmetric $\x44$ matrix
\begin{equation}
  A = \mat{
  a_{11} & a_{12} & a_{13} & a_{14} \\
  a_{12} & a_{22} & a_{23} & a_{24} \\
  a_{13} & a_{23} & a_{33} & a_{34} \\
  a_{14} & a_{24} & a_{34} & a_{44}
  },
\end{equation}
such that $a_{12},a_{14},a_{23},a_{34}\leq 0$. Therefore, $A$ is co-positive only if the following conditions are satisfied:
\begin{itemize}
  \item $a_{ii} \geq 0$, $i=1,\ldots,4$~.
  \item $a_{11}a_{22}-a_{12}^2 \geq 0$.
  \item The symmetric matrices:
  \begin{eqnarray}
    \left(
        \begin{array}{ccc}
            a_{33} \left(a_{22} a_{13}^2-2 a_{12} a_{23} a_{13}+a_{11} a_{23}^2\right) & a_{33} \left(a_{13} a_{22}-a_{12} a_{23}\right) & a_{33} \left(a_{13} a_{24}-a_{14} a_{23}\right) \\
            \cdot & a_{22} a_{33}-a_{23}^2 & a_{24} a_{33}-a_{23} a_{34} \\
            \cdot & \cdot & a_{33} a_{44}-a_{34}^2 \\
        \end{array}
    \right),\\\n
    \left(
        \begin{array}{ccc}
            a_{44} \left(a_{22} a_{14}^2-2 a_{12} a_{24} a_{14}+a_{11} a_{24}^2\right) & a_{44} \left(a_{11} a_{24}-a_{12} a_{14}\right) & a_{44} \left(a_{13} a_{24}-a_{14} a_{23}\right) \\
            \cdot & a_{11} a_{44}-a_{14}^2 & a_{13} a_{44}-a_{14} a_{34} \\
            \cdot & \cdot & a_{33} a_{44}-a_{34}^2 \\
        \end{array}
    \right),
  \end{eqnarray}
  are co-positive.
\end{itemize}
Fortunately, there is no need to review the co-positivity conditions of a $\x33$ matrix here, because the associated $\x33$ matrices of the matrix (\ref{quartic-BL-matrix}) are diagonal, hence the only condition is the non-negativity of its diagonal elements.

For a complete review of the general co-positivity conditions of any squared symmetric matrix, we suggest the Refs. \cite{Ping1993109,Hadeler198379}.
%%%%%%%%%%%%%%%%%%%%%%%%%%%%%%%%%%%%%%%%%%%%%%%%%%%%%%%%%%%%%%%%%%%%%%%%%%%%%%%%
%%%%%%%%%%%%%%%%%%%%%%%%%%%%%%%%%%%%%%%%%%
\bibliographystyle{prsty}   % this means that the order of references
                % is determined by the order in which the
                % \cite and \nocite commands appear
\bibliography{ref}  % list here all the bibliographies that
                 % you need.

\end{document}